# Design and Implementation of a Mobile Exergaming Platform

Laurent Prévost[1,2], Olivier Liechti[1] & Michael J. Lyons[2]

[1]HEIG-VD, Yverdon-les-Bains, Suisse
[2]Ritsumeikan University, Kyoto, Japan
{Laurent.Prevost, olivier.liechti}@heig-vd.ch, lyons@im.ritsumei.ac.jp

**Abstract.** This paper describes the design, implementation, and initial testing of a reusable platform for the creation of pervasive games with geo-localization services. We concentrate on role-playing games built by combining several types of simpler mini-games having three major components: Quests; Collectables; and Non-player characters (NPC). Quests encourage players to be active in their physical environment and take part in collaborative play; Collectables provide motivation; and NPCs enable player-friendly interaction with the platform. Each of these elements poses different technical requirements, which were met by implementing the gaming platform using the inTrack pervasive middle-ware being developed by our group. Several sample games were implemented and tested within the urban environment of Kyoto, Japan, using gaming clients running on mobile phones from NTT DoCoMo, Japan's largest mobile provider.

## 1  Introduction

Interactive entertainment systems traditionally offer a limited choice of user interface technologies and interaction styles that make little use of the human body and require low physical exertion. Video games, in particular, have been accused of contributing to sedentary habits and epidemic obesity amongst young people [1]. Recently, however, a new generation of video gaming devices is encouraging new styles of interactive recreation which are both more natural and involve greater physical exertion. A study at the Mayo Clinic in the USA found that children's energy expenditure increased greatly when playing a video game requiring physical activity, as compared to a traditional video game [2,3]. The spreading popularity of these new "exertion interface" games, or "exergames", is a social phenomenon of increasing benefit to the health and well being of millions of people worldwide.

In Japan, excessive interest in entertainment media, including video games, is a contributing factor to social phenomena such as the *otaku* [4] and *hikikomori* [5]. An *otaku*, (loosely "nerd" or "geek") is someone with obsessive level interest in a hobby, often involving fantasy media such as anime, manga, and games. Obsession with interactive entertainment media may complicate the socially pathological *hikikomori*

("severe social withdrawal") condition, which the Japanese Ministry of Health, Labour, and Welfare defines as individuals who isolate themselves in their homes for a period of more than six months [5].

In this work, we design and implement a platform for the development of interactive entertainment systems that encourages gaming outside the house, which is both physically and socially active. Mobile technology is reaching a stage of sophistication that can support collaborative location-aware interaction via media-rich contents. In the Japanese mobile phone market, for example, technologies such as GPS, internet connectivity, and touch sensitive high resolution colour displays have wide market penetration, offering opportunities for the development of mobile entertainment systems that area as engaging as home-based multimedia consoles. Mobile gaming platforms therefore has the potential to support a style of gaming conducive to the improvement of the physical and social health of game enthusiasts.

The development of mobile interactive entertainment systems is a relatively new area and expected to introduce new challenges for designers and engineers. For example, what genres of games will be most suitable to mobile platforms? How can game processes and data be managed in the context of mobile cooperative play? How can the (compared to home consoles) relatively limited resources of a mobile platform be most effectively used for a rich interactive entertainment experience? To explore these questions concretely we designed and implemented a geo-localized mobile gaming development platform. The work is motivated partly by the potential social benefits of mobile gaming, and partly by the interest in exploring novel forms of interaction offered by this nascent technology. The current paper is intended as a design report on lessons learned and new questions asked as a consequence of developing a working implementation of our mobile exergaming platform.

## 2 The Mobile Exergaming Platform (MEP): Game Play

For concreteness, we chose to focus this study on one style of interactive entertainment: the role-playing adventure game (RPG). This is usually a style of game requiring no physical activity, but the quests typical of RPGs should adapt well to movement in geographical space. Hence, one of the major technological objectives in the project was a generalized platform to provide the mechanisms for implementing mobile games adapted to specific locations and contexts. The platform should support tasks encouraging active exploration of the physical environment. Tasks should be general to support flexibility in site-specific implementation as well as the potential for extension and creative modification in the longer term. In this section we introduce the MEP conceptual architecture: introducing the primary components: players, non-player characters (NPC), quests, dialogs, an inventory of virtual objects, puzzles, and location data.

In keeping with the proposed mobile adventure RPG genre, tasks take the form of "quests" involving single players acting individually or multiple players in cooperative activity. Quests possess general features that were decided from the outset of the core MEP conceptual architecture design. For example, to obtain a new quest, a player uses a MEP dialog mechanism to communicate with non-player characters (NPC), a typical feature used by role-playing games to provide the narrative content of a game. Additionally, as is typical for role-playing games, MEP quests involve virtual items to be sought, collected, carried, dropped, given, traded and so on. Such items are essentially plot devices useful in motivating the player to engage in physically active, and sometimes socially cooperative play. Entertaining engagement with the physical, cultural, and social environment remains the primary goal of the mobile MEP, and the virtual objects serve to scaffold this engagement. Specific attributes of objects are undetermined in the core MEP architecture. It is intended that localized implementations of a MEP-based game will adapt objects to specific features of the geographical and cultural environs. For example, one quest might be to collect virtual flowers in the physical setting of a field or garden. To give another example, if the game is to be played in a market, a specified quest might require the player(s) to find, collect, and trade virtual foodstuffs or other goods.

Figure 1 shows the main mechanisms and entities of the game. Players, Items and Non-Player Characters are all derived from the generic entity at the basis of the MEP framework. This entity is normally associated with geographical localization data. For real players, the geo data corresponds to the player's location in physical space. For virtual entities such as items and NPCs, this is not constrained by physical reality.

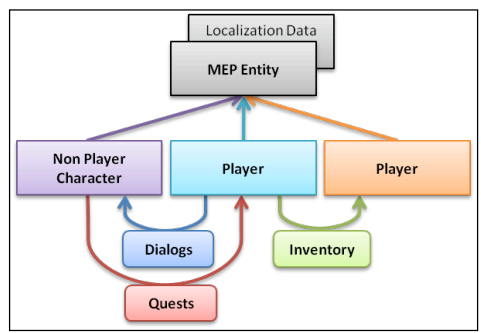

Fig. 1 – MEP Game Conceptual Architecture

In keeping with our primary target of creatively augmenting the engagement of a player with their geographical surroundings, game play primarily consists of obtaining and completing tasks with a spatial aspect. The quest mechanism is also the primary means by which players interact with the game system as well as with other players. With the current implementation, for example, quests are obtained through interactive dialog with non-player characters. The quest mechanism is highly configurable, permitting the construction of a rich variety of tasks. For concreteness we have restricted the current implementation to three major types of quests:

- Reaching a target location
- Collecting an item (virtual object)
- Solving a rebus (picture puzzle).

The MEP mechanisms allow the association of virtual objects with real-world geography. This naturally suggests quests in which a player must move through a physical environment to search for, discover, and collect such items, encouraging the player to be physically active in the exploration of an environment. Items can be placed anywhere and arranged to form a path, or placed surrounding a building or geographical feature. The MEP architecture does not restrict the positioning of items, so that the quest designer may freely use their imagination.

The third type of quest we have studied is different from the others in that it explicitly requires interaction between players. One of the design questions we are interested in is how to provide a flexible way to scaffold interaction between players. We also asked whether such interaction could be verified without having to explicitly rely on geographical data, because, for various reasons, this data sometimes has variable accuracy or is not continuously available. We satisfied both of these design criteria with the quest described next. The key idea is to provide two users with a puzzle that can only be solved by comparing data they have each obtained from the game server. We implemented a solution using a popular children's puzzle where two different pictures are supplied to two players. The players obtain these pictures from non-player characters but then must locate another (real) player to complete the puzzle. When the two pictures are viewed together in the correct sequence, the solution may be found. This type of puzzle is known as a rebus, a class of puzzle that uses pictures to represent words or parts of words [6].

## 3 MEP Architecture

This section gives a technical description of the architecture of the system used to support the game design described in Chapter 2. The MEP Architecture is based on a multi-tier application (figure 2). The main application is composed of the two sub-systems: i) inTrack, which manages localization information and ii) the game engine, which manages game information.

The inTrack platform is being developed at the University of Engineering and Management of Western Switzerland. It integrates a middleware layer that offers mechanisms for the implementation of rich location-based applications. Two APIs are exposed by inTrack. The first one is used to send localization updates for various tracking technologies. The second one is used to retrieve the state of mobile entities via various types of queries. MEP uses both APIs, since the phone itself is used to track the players.

The game server manages the state of the game entities and implements the mechanisms to support quests, dialogs, and so on. It is important to bear in mind that the game server has to manage the data link between the inTrack platform and MEP. The inTrack platform provides an API needed for this linkage of the mobile entities (viewed from inTrack) and the game entities (viewed from MEP).

In the implementation studied here, the game server also provides the client user interface running via the player's web browser. This interface is specific to the iMode technology provided by NTT DoCoMo in Japan. iMode supports use of the Internet from a DoCoMo mobile phone [7].

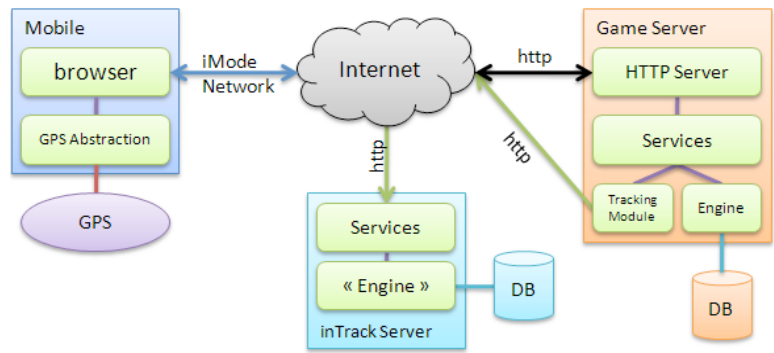

Fig. 2 – MEP multi-tier application.

Communications between the mobile client and the game server, or the game server and inTrack server take place via the HTTP protocol, integrating the diverse technologies involved in a specific MEP implementation. The game server embeds a configuration mechanism that allows building specific games based on the various mechanisms presented above. This is the main approach we have taken for adapting the platform for a specific gaming context. inTrack local mechanisms are directly integrated inside the game server application. This part of application ensures communication between the game server and inTrack server.

The mobile client in this project runs on a mobile phone from NTT DoCoMo, one of the major mobile service providers in Japan. NTT DoCoMo was selected not only for it's large share of the mobile market, but because the development environment provided by DoCoMo, a proprietary version of MIDP known as DoJa [8], seems to be more open for independent developers and better documented than those offered by the other major mobile providers.

Development of location aware applications under DoCoMo is possible by making use an attribute (known as "lcs") of the HTML tags A HREF and FORM that allows the developer to automatically add the local GPS data to the HTTP request. When this attribute is set, a dialog box appears with each HTTP request querying the user for agreement (or refusal) to send their location data. iMode-based interaction offers increased flexibility, compatibility and general ease of implementation of the game client.

## 4   User Experience with a Trial MEP Game

To evaluate MEP as a reusable platform, a concrete game was implemented in the context of the urban environment of Kyoto, Japan. This game is named the Kyoto Mobile Exergaming Project or KMEP.  Figure 3 illustrates one of the KMEP game quests, "River of Flowers". This quest requires the player to collect some flowers while walking, running, or cycling along the Kamogawa (Kamo River), a popular location for leisure activities. The quest is received from an NPC, and then starting in the South, and walking northward, one searches for and collects virtual flowers.

The "River of Flowers" quest is of the second type described in section 2. After collecting a number of flowers, the player meets another NPC, who informs the player of the status of quest completion. "River of Flowers" is a single-player quest. It implicitly requires the player to walk or ride a bicycle, in order to collect virtual flowers. While this quest is primarily a demonstration of the MEP and inTrack platforms, it also illustrates the point that the gaming experience gains the richness of the physical environment that is not obtained with a desktop gaming platform. For example, during the first run of this game, our gamer was surprised by the unexpected sight of migrating birds – a memorable event he would never have experienced indoors at home with his computer.

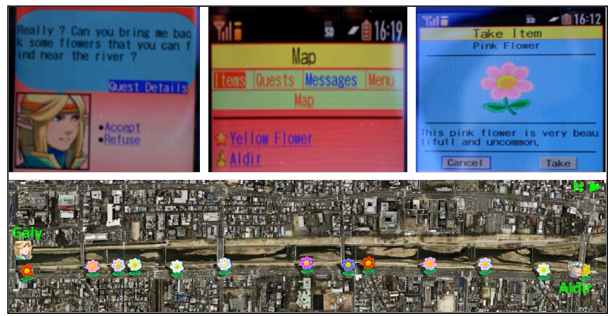

Fig. 3 – The "River of Flowers" quest.

In the Rebus category of quests, players compare pictures to solve a word puzzle. When the players obtain a solution they communicate with the MEP server for verification, which leads to the next stage of the game – for example they may then meet a new NPC and be given a quest. Rebus quests are not limited to two players. Indeed the complexity and challenge of finding the correct phrase for a rebus puzzle increases as the number of players increases.

## 5 Related Work

In this section we discuss the relationship of the current project to several prior works. The literature of ubiquitous and mobile computing is enormous, so we mention only key works that directly inspired KMEP.

City Explorer [9] is a game based on the board game Carcassonne. In City Explorer, players put some virtual tags in a real urban environment, with location categories such as bars, restaurants, and so on. Players must take a picture of the location where he has placed a virtual tag to prove he has actually visited that location. The others players can accept or refuse a player's virtual tag, based after examining the photo posted by the player. Hence localization is ensured both using GPS data and verification by the community of gamers, giving gamers an active role in validating localizations. This both authenticates physical movement in the urban environment (as opposed to purely virtual play) and introduces a social element to game play. In this project, the localization is ensured by the GPS and the acceptation of the players. The players have an active role to decide if the localizations are valid or not. In contrast to the quest-oriented RPG presented in the current paper (KMEP), the time frame of the City Explorer game is even more loosely specified and may last from one day to several weeks or longer. The game board is also not clearly defined before the game begins, but remains flexible and depends on the will of the players. By contrast, KMEP offers somewhat more structured game play, with quests being determined ahead of time by designers.

Another direct influence is the uBike project by one of the co-authors (OL) of the current paper. uBike also attempts to increase the motivation of people to do exercise [10]. In the uBike project it was found that supporting a social aspect of exercise by fostering encounters and interactions in the real world, was found to be a promising strategy for motivating exercise [11].

## 6 Conclusion

In this work we have provided a novel framework for the implementation of mobile role-playing games. We used the inTrack mobile middleware being developed at HEIG-VD to devise a new mobile exergaming platform (MEP). MEP is now fully operational and one of the positive outcomes of the work reported in this paper is the testing of this platform via its successful deployment in the local context of Kyoto, Japan.

Configuration of specific games was found to be a complex issue requiring the careful attention of the designer. To manage this complexity, game configuration was specified using a suite of XML files. This allowed a simple and flexible way to specialize the MEP for specific contexts. With its three major gaming components working solidly, the MEP allows designers to construct a variety of narrative quests, by mixing and matching quest types, collectables, and non-player characters, with no intrinsic limitations on how the types may be combined and configured.

Correctly designed, quests can offer a gaming experience conducive to physical exercise. This depends primarily on how the quest designer makes imaginative use of the geographical and cultural qualities of the context in which a game takes place. One of the quests implemented with in KMEP required a 4 km walk or bicycle ride, taking in several different terrains including shopping zones and riverside paths, leading to a fortuitous encounter with flocks of migrating birds. We believe that such games hold promise as a possible bridge for the otaku, stay-at-home technology addicts, to gently re-introduce them to the beauty and pleasure of the physical environment.

Early in this project, serious limitations of the Japanese mobile computing development environment were encountered: namely - that all Japanese providers restrict direct access to localization information of GPS enabled handsets, preventing implementation of games in which the MEP continually polled players location. This presented interesting challenges to our development process and led to some novel solutions. For example, we devised a novel method to authenticate player cooperation without building specific technical services. The Rebus puzzles, requiring real-world cooperation of multiple players, proved to be an effective way to extend existing MEP functionality and verify collaborative play.

Several aspects of MEP are points for our continuing efforts. First of all, a user-friendly authoring tool is needed to allow configuration and managing games by designers without having to edit XML files – an error-prone process which requires too much detailed knowledge of game data. Such a tool would also, ideally, provide checks for the integrity of game designs and other design support tools. We are plan to extend support for interactions between players to increase possibilities for social gaming. The current MEP architecture is sufficiently flexible to allow other types of interaction, so that research can proceed without a major revision of the platform described in the current paper. Lastly, we would like to port the user interface to other mobile providers.

## 7 Acknowledgement

This work was supported by research grants from the Japan Society for the Promotion of Science, HEIG-VD and the Hasler foundation.